\documentclass[%
 reprint,
superscriptaddress,
 amsmath,amssymb,
pre,
]{revtex4-2}

\usepackage{graphicx}% Include figure files
\usepackage{dcolumn}% Align table columns on decimal point
\usepackage{bm}% bold math
\usepackage{makecell}  %
\usepackage{booktabs}
\usepackage{color}
\usepackage{changes}
\begin{document}

%\preprint{APS/123-QED}

\title{Steady regime of radiation pressure acceleration with foil thickness adjustable within micrometers under 10-100 PW laser}% Force line breaks with \\

\affiliation{Beijing National Laboratory for Condensed Matter Physics, Institute of Physics, CAS, Beijing 100190, China}
\affiliation{Department of Physics and Beijing Key Laboratory of Opto-electronic Functional Materials and Micro-nano Devices, Renmin University of China, Beijing 100872, China}
\affiliation{Department of Mathematics and Physics, Noth China Electric Power university, Baoding, Hebei 071003, China}
\affiliation{School of Physical Sciences, University of Chinese Academy of Sciences, Beijing 100049, China}
\affiliation{Songshan Lake Materials Laboratory, Dongguan, Guangdong 523808, China}
\affiliation{Key Laboratory of Quantum State Construction and Manipulation (Ministry of Education), Renmin University of China, Beijing, 100872, China}
\affiliation{IFSA Collaborative Innovation Center, Shanghai Jiao Tong University, Shanghai 200240, China}

\author{Meng Liu}
\affiliation{Beijing National Laboratory for Condensed Matter Physics, Institute of Physics, CAS, Beijing 100190, China}
\affiliation{Department of Mathematics and Physics, Noth China Electric Power university, Baoding, Hebei 071003, China}

\author{Wei-Min Wang}%
\email{weiminwang1@ruc.edu.cn}
\affiliation{Department of Physics and Beijing Key Laboratory of Opto-electronic Functional Materials and Micro-nano Devices, Renmin University of China, Beijing 100872, China}
\affiliation{Key Laboratory of Quantum State Construction and Manipulation (Ministry of Education), Renmin University of China, Beijing, 100872, China}
\affiliation{IFSA Collaborative Innovation Center, Shanghai Jiao Tong University, Shanghai 200240, China}

\author{Yu-Tong Li}
\email{ytli@iphy.ac.cn}
\affiliation{Beijing National Laboratory for Condensed Matter Physics, Institute of Physics, CAS, Beijing 100190, China}
\affiliation{School of Physical Sciences, University of Chinese Academy of Sciences, Beijing 100049, China}

\affiliation{Songshan Lake Materials Laboratory, Dongguan, Guangdong 523808, China}
\affiliation{IFSA Collaborative Innovation Center, Shanghai Jiao Tong University, Shanghai 200240, China}

\date{\today}% It is always \today, today,

\begin{abstract}
Quasi-monoenergetic GeV-scale protons are predicted to efficiently generate via radiation pressure acceleration (RPA)  when the foil thickness is matched with the laser intensity, e.g., $L_{mat}$ at several nm to 100 nm with $10^{19}-10^{22} \rm ~W cm^{-2}$ available in laboratory. However, non-monoenergetic protons with much lower energies than prediction were usually observed in RPA experiments, because of too small foil thickness which is hard to bear insufficient laser contrast and foil surface roughness. Besides the technical problems, we here find that there is an upper-limit thickness $L_{up}$ derived from the requirement that the laser energy density should dominate over the ion source, and $L_{up}$ is lower than $ L_{mat}$ with the intensity below $10^{22} \rm~ W cm^{-2}$, which causes inefficient or unsteady RPA. As the intensity is enhanced to $\geq 10^{23} \rm ~W cm^{-2}$ provided by 10-100 PW laser facilities, $L_{up}$ can significantly exceed $L_{mat}$ and therefore RPA becomes efficient. In this regime, $L_{mat}$ acts as a lower-limit thickness for efficient RPA, so the matching thickness can be extended to a continuous range from $L_{mat}$ to $L_{up}$; the range can reach micrometers, within which foil thickness is adjustable. This makes RPA  steady and meanwhile the above technical problems can be overcome. Particle-in-cell simulation shows that multi-GeV quasi-monoenergetic proton beams can be steadily generated and the fluctuation of the energy peaks and the energy conversation efficiency remains stable although the thickness is taken in a larger range with increasing intensity. This work predicts that near future RPA experiments with 10-100 PW facilities will enter a new regime with the adjustable and large-range foil thickness for steady acceleration. 
\end{abstract}

\maketitle

%\tableofcontents

\section{Introduction}
Laser plasma interaction can provide approaches to realize compact ion accelerations due to high acceleration gradient \cite{Daido2012,Macchi2013,Fuchs2005}. The achieved ion beams with short bunch duration, compact size, and high density can be applied in fundamental science, plasma diagnostics and medical \cite{Daido2012,Macchi2013,RothPRL2001}. As one of the most attracting applications, tumor therapy 
\cite{Linz2007,Bulanov2002} demands proton beams with energy above 200 MeV and energy spread below 1$\%$\cite{Daido2012,Macchi2013}. Varieties of ion acceleration schemes have been proposed with the advancements in both high-power laser technology and targetry in the past two decades \cite{Snavely2000,WagnerPRL2016,DoverPRL2020,LiHPLSE2022,Haberberger2012,JiLLPRL,Macchi2010,BulanovRPL2015,Higginson2018}.
Among them, target normal sheath acceleration (TNSA)\cite{WilksPoP2001} is the predominant mechanism in most experiments of ion acceleration. TNSA demonstrated the cut-off proton energy near 100 MeV \cite{Higginson2018,WagnerPRL2016}, but the corresponding spectra are usually broad and the number at the cut-off energy is small. Radiation pressure acceleration (RPA) \cite{EsirkepovPRL2004,MacchiPRL2005,RobisonNJP2008,LS_Yan} is predicted to generate high-energy quasi-monoenergetic ion beam with a sufficient number in the monoenergetic peak, which has potential to meet the demands in the key applications mentioned above.

However, RPA experiments usually achieved non-monoenergetic proton beams or quasi-monoenergetic peaks at much lower energies than theoretic predictions\cite{JungPRL2011,Dollar2012,HenigPRL2009}. In RPA, the radiation pressure of an intense circularly-polarized (CP) laser pulse can push a substantial number of electrons forward, resulting in a strong charge-separation field for ion acceleration. When the radiation pressure is balanced with the charge-separation force, continuous ion acceleration can be obtained, which presents a matched foil thickness \cite{Macchi2013,LS_Yan}:
\begin{equation}
L_{mat}\simeq \frac{a_0 n_c \lambda}{\pi n_e}, \label{RPA-LS}
\end{equation}
where  $a_0= (I_0 \lambda^2/2.74\times 10^{18} \rm~ Wcm^{-2}\mu m^2)^{1/2}$ is the normalized laser amplitude, $n_c=m_e\omega^2/4\pi e^2$ is the critical density, and $I_0$, $\omega$ and $\lambda$ are the laser intensity, frequency, and wavelength, respectively. For $10^{19}-10^{22} \rm ~W cm^{-2}$ used in the existing RPA experiments \cite{AlejoPRL2022,Kuramitsu,Kim2016,HenigPRL2009,KarPRL2012,MaRPL2019}, $L_{mat}$ is at a few nm to 100 nm. With such small thickness, the foil is easy to be deformed or broken by the amplified spontaneous emission (ASE) and prepulse of the high-power laser pulse\cite{LundhRRE2007,ThauryNP2007} before the main pulse interactions with the foil. Furthermore, according to the present target fabrication technology, the surface roughness is typically at the same order with such thickness. These limitations in the current target and laser technology tend to result in inefficiency and unsteadiness of RPA, which could become worse due to transverse instabilities \cite{PegoraroPRL2007,PalmerRPL2012,WanRPL2020} and  plasma heating \cite{Paradkar2016}.

Besides the above factors, we find here that there is an upper-limit thickness $L_{up}$ for efficient RPA and $L_{up}$ is lower than or around $L_{mat}$ with the intensity below $10^{22} \rm~ W cm^{-2}$ ($a_0=61$) adopted in reported experiments\cite{Kim2016,HenigPRL2009,KarPRL2012,MaRPL2019}, which could cause inefficient or unsteady acceleration. As the intensity is enhanced to $10^{23} \rm~ W cm^{-2}$ available recently\cite{YoonOptica2021}, $L_{up}$ starts to significantly exceed $L_{mat}$, resulting in that the matching thickness can be extended to a large range from $L_{mat}$ to $L_{up}$ and up to several micrometers. This can bring both efficient and steady RPA and meanwhile the above technical problems can be overcome. In this regime, multi-GeV quasi-monoenergetic proton beams can be steadily generated and the fluctuation of the energy peaks and the energy conversation efficiency remains stable although the thickness is taken in a larger range with increasing intensity.

\section{The upper-limit thickness for steady RPA}
For the efficient and steady RPA, the requirement that the driving laser energy density should dominate over the ion source within the laser focal spot gives the upper-limit thickness:
\begin{equation}
L_{up} \le \frac{W_R\sqrt{1-v_i^2/c^2}}{\pi R^2n_im_i c^2} \label{thickness},
\end{equation}  
so that the laser pulse has enough surplus energy to transform to the ions kinetic energy, independent of the acceleration process. Here $W_R$ is the laser energy within the focal spot radius $R$, $n_i$ is the initial ion density, and $m_ic^2$ is the ion rest energy. We estimate the ion velocity $v_i$ as $v_g$ or $2v_p/(1+v_{p}^2)$, where  $v_g$ is the relativistic group velociy of laser \cite{BulanovRPL2015,Weng2012}, $v_{p}=\sqrt{\Pi}/(1+\sqrt{\Pi})$ is the piston velocity in the relativistic case \cite{EsirkepovPRL2004,Weng2012,GrechNJP2011,RobinsonPPCF2009}, and $\Pi=2a_0^2Zn_cm_e/(An_em_i)$, $Z/A$ is the charge-mass ratio, and $m_i$ is the ion mass. Our simulation below will show that they are two typical velocities in "hole-boring" \cite{PukhovPRL1997,MacchiPRL2005} and "light-sail" \cite{Macchi2010,BulanovRPL2015,LS_Qiao,LS_Chen,LS_Yu} phases in ion acceleration, respectively. 

\begin{figure}[htp]
	\includegraphics[width=0.5\textwidth, viewport=30 5 490 530, clip]{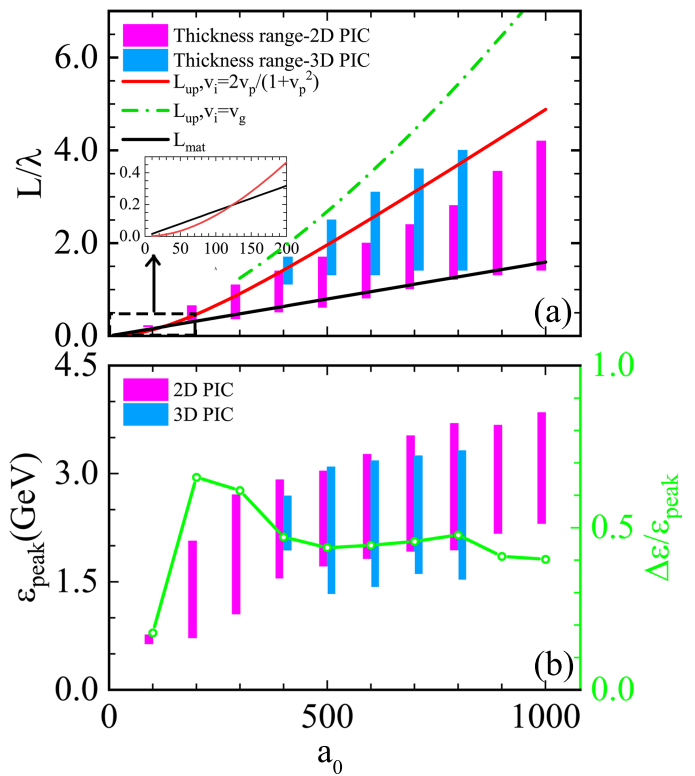}% Here is how to import EPS art
	\caption{\label{fig:Theory} (a) The target thickness $L$ for efficient RPA as a function of the laser amplitude $a_0$, where the pink-splines and blue-splines correspond 2D and 3D PIC results, respectively, the black line is $L_{mat}$ calculated from Eq.(\ref{RPA-LS}), and the red and green-dashed lines show $L_{up}$ calculated from Eq.(\ref{thickness}) with $v_i$ estimated as $2v_p/(1+v_{p}^2)$ and $v_g$, respectively. The target thickness for efficient RPA is counted when a quasi-monoenergetic proton beam is generated in PIC simulation. (b) The corresponding energy peaks $\varepsilon_{peak}$ of the quasi-monoenergetic proton beams are displayed by pink-splines and blue-splines for 2D and 3D PIC results, respectively. The green line with circles is the fluctuation of the peak energies $\Delta \varepsilon/\varepsilon_{peak}$ obtained from the 2D-PIC simulations.}
\end{figure}
According to Eqs. (\ref{RPA-LS}) and (\ref{thickness}) with a given density $n_e=200n_c$,  $L_{up}=L_{mat}=0.196 \lambda$ when $a_0=122$ (corresponding to $4\times 10^{22} \rm~ W cm^{-2}$); and $L_{up}>L_{mat}$ always holds for $a_0>122$, as shown in Fig. 1(a). Note that the intensities below and far below $4\times 10^{22} \rm~ W cm^{-2}$ were adopted in existing PRA experiments\cite{MaRPL2019,Kim2016,HenigPRL2009}. Therefore, unsteady experimental results are not only because of the too small thickness $L_{mat}$ with low tolerance to the insufficient laser contrast and foil surface roughness, but also because of the requirement that the driving laser energy density should dominate over the ion source, i.e., the thickness $L$ should be less than $L_{up}$. Adopting $L$ as $L_{mat}$ for efficient RPA has been widely recognized, so the requirement of $L_{up}>L$ is roughly equivalent to $L_{up}>L_{mat}$. As the laser intensity is higher than $4\times 10^{22} \rm~ W cm^{-2}$,  an efficient RPA with $L_{up}>L_{mat}$ starts to be possible. Further, to achieve a steady RPA, $L_{up}$ should be much greater than $L_{mat}$ and then the thickness can be chosen in a large range. For example, when $a_0=300$ (corresponding to $2.47\times 10^{23} \rm~ W cm^{-2}$), $L_{up}=0.91 \lambda$ and $L_{mat}=0.48 \lambda$ and in principle, the thickness can be taken in a range from $0.48 \lambda$ to $0.91 \lambda$. For higher laser intensity, the thickness range $\Delta L = L_{up}-L_{mat}$  is enlarged further, which can be observed in Fig.~\ref{fig:Theory}(a) and also explained in the following. One can easily derive $L_{up}\propto \xi^2\sqrt{1+2\alpha_0\xi}/(1+\alpha_0\xi)$ and $L_{mat}\propto \xi \sqrt{n_c/n_e}$, where $\alpha_0=\sqrt{2}Zm_e/(Am_i)$ and $\xi=a_0\sqrt{n_c/n_e}$ (see Supplementary Eqs.~(S5) and (S8)). For a given $n_e$ or foil species, $L_{up}$ increases more quickly than $L_{mat}$ with the growth of $a_0$, i.e., the thickness range $\Delta L$ is enlarged continuously. These analytical results are verified by our particle-in-cell (PIC) simulation results shown by the pink- and blue-splines in Fig. 1(a).

\section{PIC Simulation Results}
We perform two-dimensional (2D) and three-dimensional (3D) PIC simulations by the EPOCH code \cite{Arber}. 
A CP laser pulse with a wavelength $\lambda=1\mu m$ and an intensity profile of $I_0 exp(-2 t^2/\tau^2) exp(-y^4/R^4)$ is incident along the x direction, where the spot radius is $R=6 \lambda$ and the duration is 30 fs. The pulse arrives at the vacuum-foil interface $x=15\lambda$ at $t=0$. The foil is composed of protons $H^+$ and $e^-$ with $n_e=200n_c$.  We take a simulation box $40\lambda\times50\lambda$ ($4000\times2500$ cells in $x\times y$) moving along the x direction at the speed of light and each cell has 100 macro-particles in foil region.

Fig.~\ref{fig:Theory}(a) shows the target thickness for efficient RPA as a function of $a_0$, where the pink-splines and blue-splines correspond 2D and 3D PIC results, respectively. For a given $a_0$, we change the foil thickness and count the thickness value with which a quasi-monoenergetic GeV proton beam is generated. The counted values are illustrated by a spline representing an adjustable thickness range. One can see that the range is enlarged with the growth of $a_0$ and the pink-splines fall well between the black and red lines, in good agreement with Eq.~(\ref{RPA-LS}) and Eq.~(\ref{thickness}). This suggests that there is indeed an upper-limit thickness for efficient RPA, set by the requirement of the driver energy density dominating over the source energy density.

The PIC results also indicate that: the well-known matching thickness $L_{mat}$ acts as a lower-limit value for efficient RPA, and then the matching thickness originally as an isolated value point can be extended to a continuous range. This is because $L_{mat}$ is derived under an ideal condition that the foil electrons as a whole is pushed forward and form charge-separation field to balance with the laser radiation pressure exerted on the electrons. Actually, only part of the foil electrons can be pushed forward out of the foil, which is becomes more significant for a relatively large thickness with high laser intensity. Furthermore, the electrostatic force of charge-separation field $E_x=4\pi en_eL$ should be higher than the radiation pressure $2I/c$. Otherwise, the foil electrons can be blown out, the compressed electron layer cannot be formed, and ions cannot be accelerated. Hence,
\begin{equation}
L\geq L_{mat},
\end{equation}  
should be a more reasonable condition for sustaining the charge-separation field to accelerate ions continuously.  It should be noted that Eqs. (\ref{RPA-LS}) and (\ref{thickness}) are given in the 1D case. The omitted transverse effects tend to deteriorate the target and a thicker foil is needed to overcome the deterioration. Thus, 2D PIC results are in better agreement with $L_{mat}$ and $L_{up}$ than 3D ones.

The enlarged efficient thickness range bounded by $L_{mat}$ and $L_{up}$ provides a favorable freedom for foil thickness choice and the matched thickness can be adopted as a value much higher than the original prediction by $L_{mat}$. For instance, quasi-monoenergetic GeV proton beams can be stably generated from the foil with a thickness within $0.4\lambda$ - $1.2\lambda$ for $a_0=300$, and $0.8\lambda$ - $2.1\lambda$ for $a_0=600$ from PIC results (also see Supplementary Note 1 and Fig.~S2). When $a_0\geq1000$, the thickness range $\Delta L$ even enlarges above $2\lambda$, favoring the target design in future experiments.

Although the thickness is taken in a larger range with increasing $a_0$, the fluctuation of the energy peaks remains stable, as shown in Fig.~\ref{fig:Theory}(b). This figure plots the energy peaks $\varepsilon_{peak}$ of the quasi-monoenergetic proton beams obtained from 2D and 3D PIC results. In the typical simulation with $a_0=300$, the peak energy decreases from 3.6 GeV to 1.2 GeV as the foil thickness increases from $0.4\lambda$ to $1.2\lambda$. With a larger amplitude $a_0=600$, the peak energy only decreases from 4.5 GeV to 3.0 GeV as the foil thickness increases from $0.8\lambda$ to $2.1\lambda$ (also see Supplementary Note 1 and Fig.~S2). The green line in Fig.~\ref{fig:Theory}(b) displays the energy fluctuation $\Delta \varepsilon/\varepsilon_{peak}$ as a function of $a_0$. It is shown that the fluctuation reaches $70\%$ at $a_0=200$, decreases to $40\%$ at $a_0=400$, and then maintains around this value since $a_0 \ge 400$. Even the thickness range $\Delta L$ is above $2\lambda$ with $a_0\geq1000$, $\Delta \varepsilon/\varepsilon_{peak}$ does not grow. This is because the proton velocity or peak energy is mainly determined by the laser relativistic group velocity which increases slowly with the growing $a_0$ when $a_0$ is sufficiently large.  The slowly increasing group velocity also causes that the energy conversion efficiency of the protons basically remains around $20\%-25\%$ as shown in the Supplementary Note 2, Fig.~S4 and Table S1.
\begin{figure}[htp]
	\includegraphics[width=0.5\textwidth, viewport=10 5 600 420, clip]{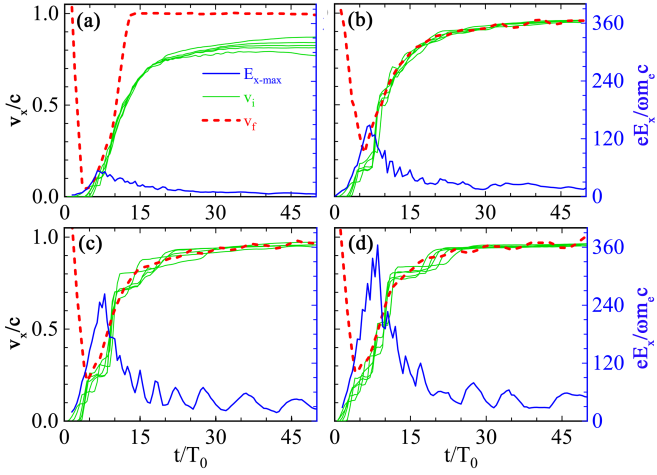}% Here is how to import EPS art
	\caption{\label{fig:velocity} Temporal evolution of the laser wavefront velocity $v_f$ (red dotted-line) and tracked proton velocity $v_i$ (green-line), where we track 100 protons gaining high energies finally and take 5 typical ones. Evolution of the maximum of the longitudinal electric field $E_{x-max}$ (corresponding to the right $y-$axis) is also displayed by the blue line. Here the laser amplitude $a_0$ and target thickness $L$ are taken as  (100, $0.18\lambda$), (300, $0.56\lambda$), (500, $1.1\lambda$) and (700, $1.7\lambda$) in (a)-(d), respectively.}
\end{figure}

Fig.~\ref{fig:velocity} shows the evolution of tracked proton velocity $v_i$ and the laser wavefront velocity $v_f$ representing the group velocity, where $v_f$ is defined as the moving velocity of the laser intensity surface $I_0/10$ and $I_0$ is the initial intensity. The evolution of $v_f$ can be separated into two stages in Fig.~\ref{fig:velocity}(b)-(d). In the first stage with $t \lesssim 11-12T_0$, $v_f$ firstly decreases dramatically at the beginning interaction of the laser with the foil, and later increases quickly along with the foil being pushed forward by laser radiation pressure. This stage has been widely studied\cite{MacchiPRL2005,RobinsonPPCF2009} and $v_i$ can be estimated as $2v_p/(1+v_{p}^2)$. At the second stage after about $12T_0$, $v_f$ becomes roughly constant and $v_i$ is very close to $v_f$ in Figs.~\ref{fig:velocity}(b)-\ref{fig:velocity}(d) with $a_0=300-700$, meaning that the protons are efficiently and continuously accelerated and then move along with the laser pulse. In this stage, $v_i$ can be estimated by the relativistic group velocity $v_g$\cite{BulanovRPL2015,PegoraroPRL2007,SVBulanovPoP}, where $v_i\simeq 0.973c$ and $v_i\simeq 0.982c$ for the cases with $a_0=500$ and $a_0=700$ read from Fig.~\ref{fig:velocity}(c) and \ref{fig:velocity}(d). In this case, the group velocity $v_g/c \simeq 1-\frac{n_e}{\sqrt{2}a_0n_c}$ grows slowly with $a_0$ ($a_0\gg 1$). This agrees with Fig. 1(b) that the peak energy increases slowly from $a_0=400$ to $a_0=1000$. By contrast, in Fig.~\ref{fig:velocity}(a) with $a_0=100$, the protons velocity can only reach $0.75c$ much lower than $v_f$ because the laser wavefront breaks through the foil and the protons cannot catch up.  

\begin{figure}[htp]
	\includegraphics[width=0.5\textwidth, viewport=5 5 580 280, clip]{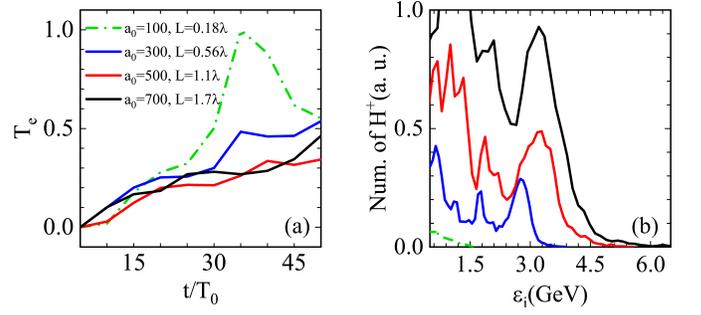}  % Here is how to import EPS art
	\caption{\label{fig:Teene} (a) Evolution of the plasma electron temperature and  (b) the energy spectra of protons at $t=70T_0$, where different lines in (a) and (b) represent different ($a_0$, $L$) corresponding to the parameters taken in Figs.~\ref{fig:velocity}(a)-\ref{fig:velocity}(d), respectively. The plasma temperature are calculated with the electrons in the compressed density layer and normalized by that in the case of $a_0=100$.}
\end{figure}

Fig.~\ref{fig:Teene} shows that the plasma heating is suppressed with the growing laser intensity, facilitating the acceleration. Compared with the case with $a_0=100$, the plasma temperature is reduced by 50\% with $a_0=300$ and 75\% with $a_0=500$ and 700 at the second stage. In efficient RPA with the high intensities, the protons and electrons move along with the laser pulse and then their velocities [see Fig.~\ref{fig:velocity}(b)-(d)] are close $c$ and mainly in the longitudinal direction, i.e., most of the particle energies are longitudinal, among which the protons cover most energies. This causes significant reductions of the temperature and the transverse spread of electrons.

\begin{figure}[htp]
	\includegraphics[width=0.7\textwidth, viewport=5 35 500 450, clip]{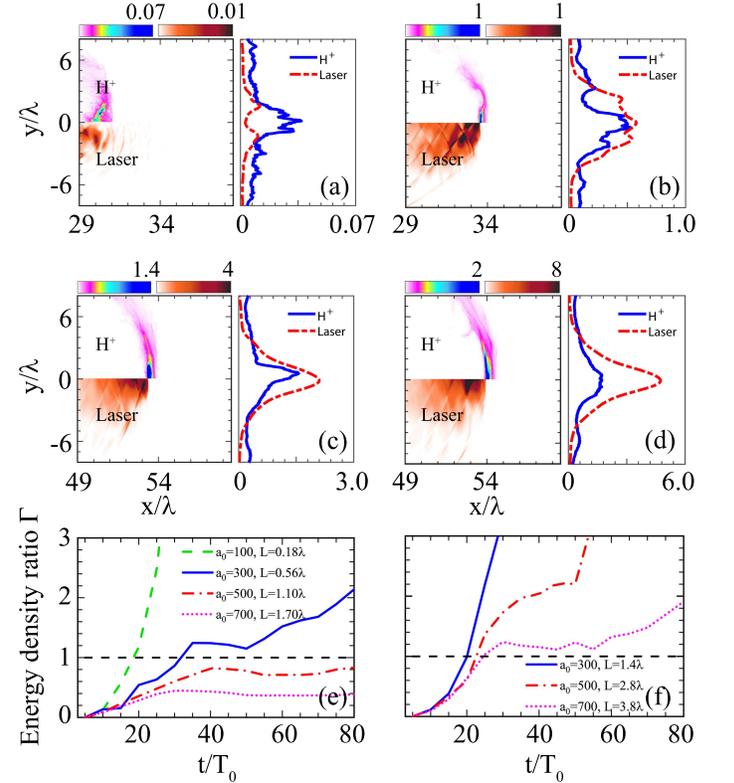}% Here is how to import EPS art
	\caption{\label{fig:enedensity} (a)-(d) Spatial distributions of energy densities of the protons (top half) and the laser (bottom half) and the energy densities integrated along $x$ versus $y$ are also plotted by curves, where (a) and (b) are the cases with $a_0=100/300$ at $t=30T_0$ and (c) and (d) are the cases with $a_0=500/700$ at $t=50T_0$. [(e), (f)] Temporal evolution of the energy density ratio between the protons and the laser, where the ratio is calculated by the energy density peaks of the protons and the laser.}
\end{figure}
The upper-limit thickness $L_{up}$ in Eq.~(\ref{thickness}) is given by the requirement that the driving laser energy density should dominate over the proton source within the laser focal spot during the efficient RPA. The requirement is verified by Fig.~\ref{fig:enedensity}, which displays the energy densities of the protons and the laser. When the thickness is in the range from $L_{mat}$ to $L_{up}$ for the efficient RPA, e.g., Figs.~\ref{fig:enedensity}(b)-(d), the laser energy density is dominant over the proton energy density. Fig.~\ref{fig:enedensity}(e) illustrates the evolution of the energy density ratio $\Gamma$ of the protons to the laser. $\Gamma$ is less than 1 in the whole simulation duration for ($a_0$, $L$) = (500, $1.1\lambda$) and (700, $1.7\lambda$)  and before $t=30T_0$ for (300, $0.56\lambda$) [also see Fig.~\ref{fig:enedensity}(b) given at $t=30T_0$], which corresponds to the efficient acceleration time. While the thickness is larger than $L_{up}$ for the given $a_0$=300, 500, and 700, shown in Fig.~\ref{fig:enedensity}(f), $\Gamma$ starts to be more than 1 as early as about $t=20T_0$ and the acceleration is inefficient (i.e., no quasi-monoenergetic peak or much lower peak energy).

\section{Conclusion and discussion}

In summary, we have found there is an upper-limit thickness $L_{up}$ for efficient RPA, deriving from the requirement that the driving laser energy density should dominate over the proton source. The well-known matching thickness $L_{mat}$ acts as a lower-limit value, and therefore, the matching thickness originally as an isolated value point can be extended to a continuous range from $L_{mat}$ to $L_{up}$.  $L_{up}>L_{mat}$ can be achieved for steady RPA with $I_0>4\times10^{22}\rm ~ W cm^{-2}$ and the thickness range $L_{up}-L_{mat}$ is enlarged with the laser intensity. For $10^{23}\sim 10^{24} \rm ~W cm^{-2}$ delivered from the 10 PW and 100 PW laser facilities\cite{Gan2021,Christophe2022,Tanaka2020MPE}, the thickness range can reach a few micrometers providing favorable freedom for foil thickness choice in RPA experiments. Although the thickness is taken in an larger range with increasing intensity, the fluctuation of the energy peaks as well as the energy conversation efficiency remains stable. This work predicts that near future RPA experiments with 10-100 PW laser facilities will enter a new regime with the adjustable and large-range target thickness for steady ion acceleration.

Note that in Eq.~(\ref{thickness}) the proton velocity $v_i$ is mainly estimated by the laser relativistic group velocity $v_g$, which has been verified by Fig.~\ref{fig:velocity}, and the laser energy is calculated with the spot radius at the focusing plane, which does not change significantly with laser propagation within the Rayleigh length (113 $\mu m$ in our case). Therefore, Eq.~(\ref{thickness}) can give a reasonable $L_{up}$ close to the PIC simulations. 

We also check the influence of the strong field quantum electrodynamics (QED)  effects \cite{Berestetskii1982,Baier1998,Ritus1985,Piazza2012} on RPA. The influence enhances with the growing $a_0$, but basically it can be negligible (see Supplementary Note 3 and Fig.~S5). With $a_0=1000$, the energy conversion efficiency of the gamma-photons increases with the target thickness and it reaches $8\%$ at the maximum thickness $4.4 \lambda$ for efficient RPA, where both the energy peak and energy conversion of the protons are reduced by less than $6\%$. In efficient RPA, electrons move mainly along the laser propagating direction which makes small QED parameters and weak QED effects \cite{Piazza2012,TamburiniNJP2010}. 

Besides the foil species composed of $n_H=n_e=200n_c$, we also investigated the steady RPA process for the lower density with $n_H=n_e=100n_c$(see Supplementary Table S1). Moreover, the realistic targets, lithium hydride(LiH)\cite{DavisPOP2009} with different thicknesses, are also adopted and the resutls agree with Eq (2)(see Supplementary Note 4 and Fig.~S6).

%\nocite{*}

\begin{acknowledgments}
	This work was supported by the National Key R\&D Program of China (Grant No. 2018YFA0404801), National Natural Science Foundation of China (Grants No. 12205366), the Strategic Priority Research Program of Chinese Academy of Sciences (Grant Nos. XDA25050300, XDA25010300), and the Fundamental Research Funds for the Central Universities (Grant Nos. 2020MS138).
\end{acknowledgments}

%\nocite{*}

\end{document}